\documentclass[runningheads]{svmult}

\usepackage{makeidx}   
\usepackage{graphicx}  
\usepackage{subeqnar}  
\usepackage{multicol}  
\usepackage{physprbb}  
\makeindex             

\begin{document}
\title*{Star Clusters in the Galactic Anticenter Stellar Structure:
 New Radial Velocities \& Metallicities}
\toctitle{Star Clusters in the Galactic Anticenter Stellar Structure:
\protect\newline New Radial Velocities \& Metallicities}
%
%
\titlerunning{Star Clusters in the GASS}
%
\author{Peter M. Frinchaboy\inst{1}
\and Ricardo R. Mu\~{n}oz\inst{1}
\and Steven R. Majewski\inst{1}
\and \\ Eileen D. Friel\inst{2}
\and Randy L. Phelps\inst{2,3}
\and William B. Kunkel\inst{4}}
\authorrunning{Peter Frinchaboy et al.}
%
%

\institute{University of Virginia, P. O. Box 3818, Charlottesville, VA 22903, USA
\and National Science Foundation, 4201 Wilson Boulevard, Arlington, Virginia 22230 
\and C.~S.~U., Sacramento, 6000 J Street, Sacramento, CA 95819
\and Las Campanas Observatory, Casilla 601, La Serena, Chile}

\maketitle              

\begin{abstract}
The Galactic Anticenter Stellar Structure (GASS) has been identified
with excess surface densities of field stars in several large area sky
surveys, and with an unusual, string-like grouping of star clusters.
Some members of the cluster grouping
have radial velocities (RVs) consistent with the observed GASS velocity-longitude trend.
We provide new RV measurements of stars in six clusters
that have been suggested to be associated with the GASS. We show
that the RVs of at least four clusters are
consistent with the previously measured RV trend for GASS. We also derive
spectroscopic metallicities for four clusters, and provide an
improved age-metallicity relation for the clusters apparently associated with GASS.
\end{abstract}


We have found that the outer most open clusters found to date seem to 
be strung along the Galactic
anticenter stellar structure (GASS) \cite{fri:pmf04}.  
This structure \cite{fri:newberg}, \cite{fri:yanny}, \cite{fri:ibata}, \cite{fri:majewski},  
\cite{fri:helio}, \cite{fri:crane} was discovered as an excess of stars beyond the apparent 
limit of the Galactic disk.
Previous work \cite{fri:crane} has resulted in a number of parameters of the stream including: 
(1) a velocity-longitude trend indicating
a slightly non-circular orbit, (2) a velocity dispersion smaller than
even that of disk stars, (3) a wide metallicity spread from $-1.6 <$ [Fe/H]$ < -0.4$.

%

\begin{figure}[b]
\begin{center}
\includegraphics[width=.580\textwidth]{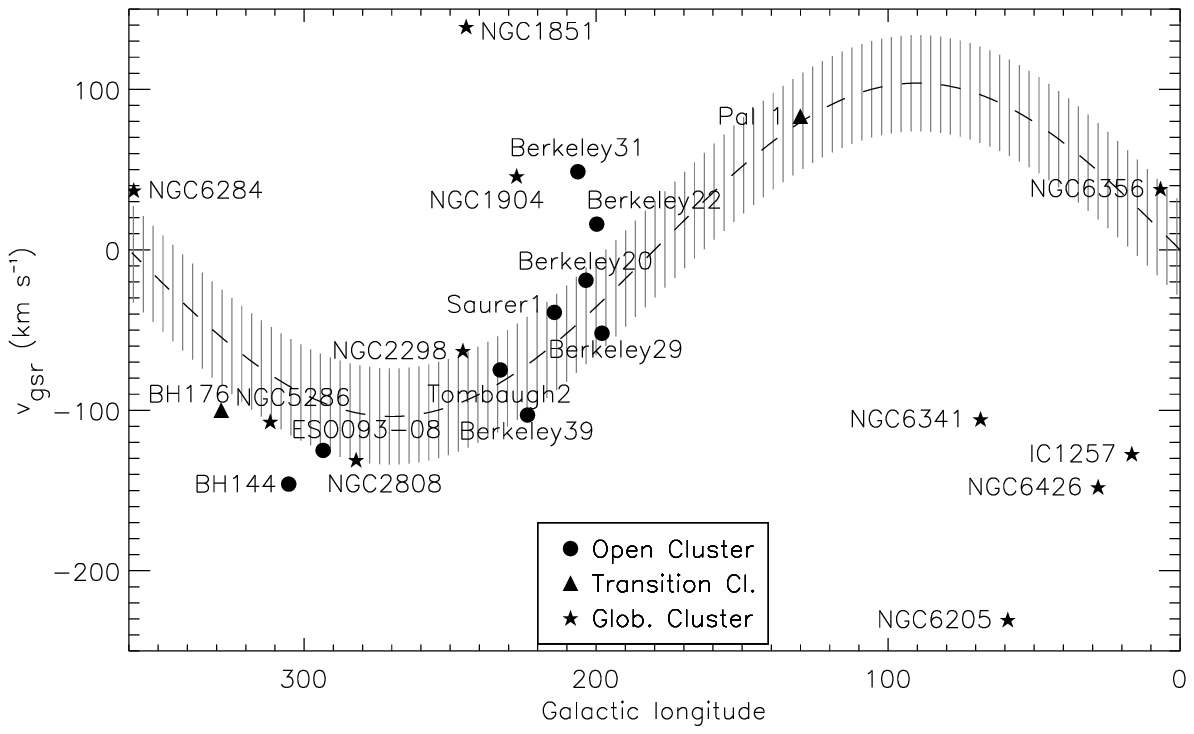} 
\includegraphics[width=.355\textwidth]{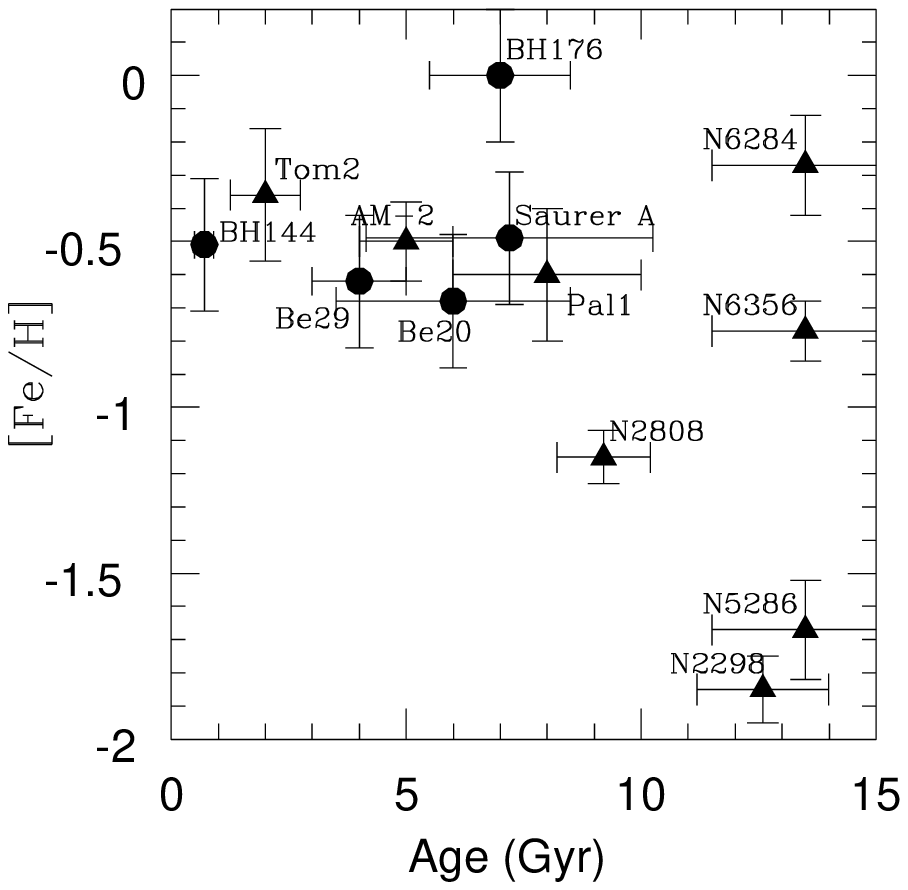} 
\end{center}
\caption[]{(a,left) The $l-v_{GSR}$ distribution of objects lying
  within 2.35 kpc of the GASS cluster plane.  
  The hash marks represent a velocity
  dispersion of 30 km s$^{-1}$ about the $v_{gsr}$ of a circularly
  orbiting object at r$_{GC}=18$ kpc with $v_{circ}=220$ km s$^{-1}$,
  which reasonably matches GASS M giant velocities\cite{fri:crane}. 
  (b,right) AMR of all GASS candidate clusters.  
Circles denote values measured and listed in table, triangles are data
from \cite{fri:pmf04}.}
\label{fri:f1}
\end{figure}

Spectra for stars in the clusters Berkeley 29, BH 176, and Saurer 1 were collected with the Hydra Spectrograph 
and clusters Berkeley 22,  BH 144(ESO 096-SC04), and ESO 093-SC08 were collected 
using the RC Spectrograph on 
the Blanco 4-meter telescope. 
The data were reduced using standard IRAF reduction methodology with
RVs determined using {\ttfamily fxcor}. 
The CaII infrared triplet was measured using the index definitions of 
\cite{fri:az}, transformed to the common system derived in \cite{fri:cole}.
%
%
The resulting RVs and metallicities are presented in Table~\ref{fri:Tab1}.  The typical error 
in the metallicities ($\sigma_{[Fe/H]}$) are $\sim$ 0.3 dex.  All clusters measured are 
shown with the GASS longitude-velocity ( $l$ vs. $V_{gsr}$) trend (Fig.~\ref{fri:f1}a) 
and members are used to produce an 
age-metallicity relation 
(Fig.~\ref{fri:f1}b). 
We find that the clusters Berkeley 29, Saurer 1, ESO 0903-SC08, and possibly BH 144 \& 176, have 
RVs 
consistent with being part of the GASS cluster 
system; however, further work is needed to confirm actual membership in GASS.

\begin{table}
\caption{GASS Cluster Candidate Radial Velocities and Metallicities        }
\begin{center}
\renewcommand{\arraystretch}{0.93}
\setlength\tabcolsep{5pt}
\begin{tabular}{lccccccc}
\hline\noalign{\smallskip}
Cluster & $l$  &  $b$  & \# stars & $V_r$       & $V_{gsr}$
  & [Fe/H] & GASS?\\[-0.5ex]
        &      &       & \scriptsize{Metallicity(RV)}& \scriptsize{(km/s)} &  \scriptsize{(km/s)}
  &        &      \\
\noalign{\smallskip}
\hline
\noalign{\smallskip}
Berkeley 22            & 199.8 &  $-$8.1  &  4(5)  & 106 $\pm$ 9  &  $+$16 & $-$0.97 & N \\
Berkeley 29            & 198.0 &  $+$8.0  &  8(8)  &  26 $\pm$ 6  &  $-$52 & $-$0.62 & Y \\
Berkeley 20            & 203.5 & $-$17.3  &  3(3)  &  78 $\pm$ 6  &  $-$19 & $-$0.68 & (Cal) \\
Saurer 1               & 214.3 &  $-$6.8  &  2(4)  &  98 $\pm$ 9  &  $-$39 & $-$0.49 & Y$^1$ \\
Berkeley 39            & 223.5 & $+$10.1  & 19(19) &  59 $\pm$ 4  & $-$103 & $-$0.27 & (Cal) \\
ESO093$-$SC08          & 293.5 &  $-$4.0  &  0(1)  &  86 $\pm$ 10 & $-$125 &  ...... & Y \\
BH 144  \scriptsize{(ESO096$-$SC04)}   
                       & 305.3 &  $-$3.2  &  2(3)  &  40 $\pm$ 10 & $-$146 & $-$0.51 & ?$^2$ \\
47 Tuc                 & 305.9 & $-$44.9  &  5(.)  &  .......     &  ..... & $-$0.78 & (Cal) \\
BH 176                 & 328.4 &  $-$4.3  &  0(3)  &  13 $\pm$ 5  & $-$100 &  ...... & Y$^2$ \\
\hline 
\end{tabular}
\end{center}
{$^1$\scriptsize 
We find Saurer 1 to be a member due to its $V_{gsr}$; \cite{fri:carraro} did not correct this RV to $V_{gsr}$ before \\[-1ex] 
incorrectly excluding Saurer 1 from the GASS member clusters.  $^2$ It is noted from the spatial \\[-1ex]
distribution of the clusters that GASS should have an elliptical orbit.  If this change is made, \\[-1ex] 
BH 144 and BH 176, along with NGC5286 and NGC 2808, should fit the corrected $l-v_{GSR}$ trend.
 \\[-7ex]} 
\label{fri:Tab1}
\end{table}

We acknowledge funding by NSF grant AST-0307851, NASA/JPL contract 1228235,
the David and Lucile Packard Foundation, Robert J. Huskey Travel fellowship, 
AAS International Travel Grant, 
and the F.H. Levinson Fund of the Peninsula Community Foundation.
%

%

\end{document}